\documentclass[twoside,twocolumn,9pt]{article}
\usepackage{extsizes}
\usepackage[super,sort&compress,comma]{natbib} 
\usepackage[version=3]{mhchem}
\usepackage[left=1.5cm, right=1.5cm, top=1.785cm, bottom=2.0cm]{geometry}
\usepackage{balance}
\usepackage{mathptmx}
\usepackage{sectsty}
\usepackage{graphicx} 
\usepackage{lastpage}
\usepackage[format=plain,justification=justified,singlelinecheck=false,font={stretch=1.125,small,sf},labelfont=bf,labelsep=space]{caption}
\usepackage{float}
\usepackage{fancyhdr}
\usepackage{fnpos}
\usepackage[english]{babel}
\addto{\captionsenglish}{%
  
}
\usepackage{array}
\usepackage{droidsans}
\usepackage{charter}
\usepackage[T1]{fontenc}
\usepackage[usenames,dvipsnames]{xcolor}
\usepackage{setspace}
\usepackage[compact]{titlesec}
\usepackage{hyperref}
\usepackage{amssymb}
\usepackage{hhline}
\usepackage{url}
\usepackage{epstopdf}%This line makes .eps figures into .pdf - please comment out if not required.
\graphicspath{{./Figs}}
\usepackage{dcolumn}% Align table columns on decimal point
\usepackage{bm}% bold math
\newcommand{\bvec}[1]{\bm{\mathrm{#1}}}
\usepackage{xcolor,soul}
\usepackage[ISO]{diffcoeff}

\newcommand{\rgl}{\rangle}

\usepackage[nameinlink]{cleveref}
\definecolor{cream}{RGB}{222,217,201}

\begin{document}

\pagestyle{fancy}
\thispagestyle{plain}
\fancypagestyle{plain}{
%%%HEADER%%%
\renewcommand{\headrulewidth}{0pt}
}
%%%END OF HEADER%%%

%%%PAGE SETUP - Please do not change any commands within this section%%%
\makeFNbottom
\makeatletter
\renewcommand\LARGE{\@setfontsize\LARGE{15pt}{17}}
\renewcommand\Large{\@setfontsize\Large{12pt}{14}}
\renewcommand\large{\@setfontsize\large{10pt}{12}}
\renewcommand\footnotesize{\@setfontsize\footnotesize{7pt}{10}}
\makeatother

\renewcommand{\thefootnote}{\fnsymbol{footnote}}
\renewcommand\footnoterule{\vspace*{1pt}% 
\color{cream}\hrule width 3.5in height 0.4pt \color{black}\vspace*{5pt}} 
\setcounter{secnumdepth}{5}

\makeatletter 
\renewcommand\@biblabel[1]{#1}            
\renewcommand\@makefntext[1]% 
{\noindent\makebox[0pt][r]{\@thefnmark\,}#1}
\makeatother 
\renewcommand{\figurename}{\small{Fig.}~}
\sectionfont{\sffamily\Large}
\subsectionfont{\normalsize}
\subsubsectionfont{\bf}
\setstretch{1.125} %In particular, please do not alter this line.
\setlength{\skip\footins}{0.8cm}
\setlength{\footnotesep}{0.25cm}
\setlength{\jot}{10pt}
\titlespacing*{\section}{0pt}{4pt}{4pt}
\titlespacing*{\subsection}{0pt}{15pt}{1pt}
%%%END OF PAGE SETUP%%%

%%%FOOTER%%%
\fancyfoot{}
\fancyfoot[LO,RE]{\vspace{-7.1pt}\includegraphics[height=9pt]{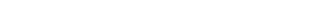}}
\fancyfoot[CO]{\vspace{-7.1pt}\hspace{13.2cm}\includegraphics{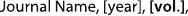}}
\fancyfoot[CE]{\vspace{-7.2pt}\hspace{-14.2cm}\includegraphics{head_foot/RF}}
\fancyfoot[RO]{\footnotesize{\sffamily{1--\pageref{LastPage} ~\textbar  \hspace{2pt}\thepage}}}
\fancyfoot[LE]{\footnotesize{\sffamily{\thepage~\textbar\hspace{3.45cm} 1--\pageref{LastPage}}}}
\fancyhead{}
\renewcommand{\headrulewidth}{0pt} 
\renewcommand{\footrulewidth}{0pt}
\setlength{\arrayrulewidth}{1pt}
\setlength{\columnsep}{6.5mm}
\setlength\bibsep{1pt}
%%%END OF FOOTER%%%

%%%FIGURE SETUP - please do not change any commands within this section%%%
\makeatletter 
\newlength{\figrulesep} 
\setlength{\figrulesep}{0.5\textfloatsep} 

\newcommand{\topfigrule}{\vspace*{-1pt}% 
\noindent{\color{cream}\rule[-\figrulesep]{\columnwidth}{1.5pt}} }

\newcommand{\botfigrule}{\vspace*{-2pt}% 
\noindent{\color{cream}\rule[\figrulesep]{\columnwidth}{1.5pt}} }

\newcommand{\dblfigrule}{\vspace*{-1pt}% 
\noindent{\color{cream}\rule[-\figrulesep]{\textwidth}{1.5pt}} }

\makeatother
%%%END OF FIGURE SETUP%%%

%%%TITLE, AUTHORS AND ABSTRACT%%%
\twocolumn[
  \begin{@twocolumnfalse}
{\includegraphics[height=30pt]{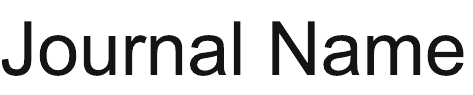}\hfill\raisebox{0pt}[0pt][0pt]{\includegraphics[height=55pt]{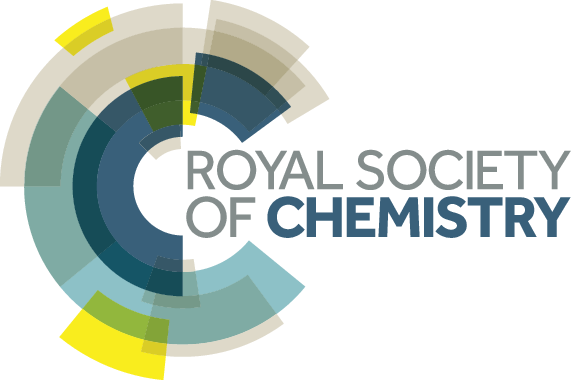}}\\[1ex]
\includegraphics[width=18.5cm]{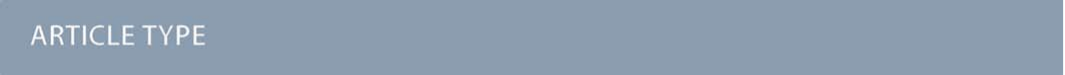}}\par
\vspace{1em}
\sffamily
\begin{tabular}{m{4.5cm} p{13.5cm} }

\includegraphics{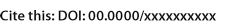} & \noindent\LARGE{\textbf{Rapidly convergent {\color{black}quantum Monte} Carlo using a Chebyshev projector}} \\%Article title goes here instead of the text "This is the title"
\vspace{0.3cm} & \vspace{0.3cm} \\

 & \noindent\large{Zijun Zhao,$^{\ast}$\textit{$^{a}$}$^\ddag$ Maria-Andreea Filip,\textit{$^{a}$}\textsuperscript{\textsection} and Alex J W Thom\textit{$^{a}$}} \\%Author names go here instead of "Full name", etc.

\includegraphics{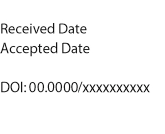} & \noindent\normalsize{
The multi-reference coupled-cluster Monte Carlo (MR-CCMC) algorithm is a determinant-based quantum Monte Carlo (QMC) algorithm that is conceptually similar to Full Configuration Interaction QMC (FCIQMC). It has been shown to offer a balanced treatment of both static and dynamic correlation while retaining polynomial scaling, although application to large systems with significant strong correlation remained impractical. In this paper, we document recent algorithmic advances that enable rapid convergence and a more black-box approach to the multi-reference problem. These include a logarithmically scaling metric-tree based excitation acceptance algorithm to search for determinants connected to the reference space at the desired excitation level and a symmetry-screening procedure for the reference space. We show that, for moderately sized reference spaces, the new search algorithm brings about an approximately 8-fold acceleration of one MR-CCMC iteration, while the symmetry screening procedure reduces the number of active reference space determinants at essentially no loss of accuracy. We also introduce a stochastic implementation of an approximate wall projector, which is the infinite imaginary time limit of the exponential projector, using a truncated expansion of the wall function in Chebyshev polynomials. {\color{black}Notably, this wall-Chebyshev projector can be used to accelerate any projector-based QMC algorithm. We show that it requires significantly fewer applications of the Hamiltonian to achieve the same statistical convergence. We benchmark these acceleration methods on the beryllium and carbon dimers, using initiator FCIQMC and MR-CCMC with basis sets up to cc-pVQZ quality.} } \\%The abstrast goes here instead of the text "The abstract should be..."

\end{tabular}

 \end{@twocolumnfalse} \vspace{0.6cm}

  ]
%%%END OF TITLE, AUTHORS AND ABSTRACT%%%

%%%FONT SETUP - please do not change any commands within this section
\renewcommand*\rmdefault{bch}\normalfont\upshape
\rmfamily
\section*{}
\vspace{-1cm}

\footnotetext{\textit{$^{a}$~Yusuf Hamied Department of Chemistry, University of Cambridge,
Cambridge, UK. E-mail: zz376@cantab.ac.uk}}
\footnotetext{\textit{$^{\ddag}$}~Present address: \textit{Department of Chemistry, Emory University, Atlanta, GA, USA. }}
\footnotetext{\textsection~Present address: \textit{Max Planck Institute for Solid State Research, Stuttgart, Germany.} }

\section{Introduction}

Quantum Monte Carlo (QMC) methods have long provided a powerful alternative to conventional electronic structure methods, by generating high accuracy results at a fraction of the cost of standard approaches. The combination of Variational Monte Carlo (VMC)\cite{McMillan1965,Ceperley1977} and Diffusion Monte Carlo (DMC)\cite{fixnode75, fixnode_anderson,Anderson1980} has become a significant benchmarking approach in many areas of electronic structure,\cite{Grossman2002,Nemec2010,Cox2014,Ganesh2014,DellaPia2022}, but it is limited by the need to provide some approximate nodal structure to avoid collapse onto bosonic solutions. Fermionic Monte Carlo methods\cite{booth_fermion_2009, thom_stochastic_2010} have since been developed which act directly in the anti-symmetrised Hilbert space of the electronic structure problem, thereby removing the potential for bosonic solutions \textit{a priori}.

First introduced in 2009 by Booth \textit{et al.}\cite{booth_fermion_2009}, full configuration interaction quantum Monte Carlo (FCIQMC) can be variously described as a stochastic power iteration algorithm or an iterative solution to the imaginary time Schr{\" o}dinger's equation. Here, we give a brief summary of theoretical underpinnings of FCIQMC by taking the latter view. By applying a Wick rotation \cite{wickPropertiesBetheSalpeterWave1954}, $\tau\leftarrow it,\ \tau\in\mathbb{R}$, to the time-dependent Schr{\" o}dinger's equation ${i\dot{\Psi}=\hat{H}\Psi}$, one obtains the imaginary time Schr{\" o}dinger's equation:
\begin{equation}
    \diffp{}{\tau}|\Psi(\tau)\rangle=-\hat{H}|\Psi(\tau)\rangle,\ \tau\in\mathbb{R}.
\end{equation}
It can be formally integrated to give
\begin{equation}
\label{eq:projector}
    |\Psi(\tau)\rangle=e^{-\tau(\hat{H}-S)}|\Psi(0)\rangle,
\end{equation}
with $S$ being the constant of integration, also known as the `shift'.

The reference wavefunction $|\Psi(0)\rangle$, commonly a Hartree--Fock (HF) solution, can be expanded in the eigenbasis of the full Hamiltonian,  $\{|\Psi_i^{\mathrm{FCI}}\rangle\}$, leading to
\begin{equation}
    |\Psi(\tau)\rangle=\sum_ie^{-\tau(E_i-S)}c_i|\Psi_i^{\mathrm{FCI}}\rangle,
\end{equation}
with $\{E_i\}$ being the eigenspectrum of the full Hamiltonian and $|\Psi(0)\rangle = \sum_i c_i |\Psi_i^{\mathrm{FCI}}\rangle$. We can see that, if $S=E_0$, in the limit of $\tau\rightarrow\infty$, we obtain the ground state of the full Hamiltonian. 

By discretising the projector in \Cref{eq:projector} and further applying the first-order Taylor series expansion, we obtain the `master equation' of FCIQMC:
\begin{equation}
\label{eq:linear_proj}
    |\Psi(\tau+\delta\tau)\rangle=[1-\delta\tau(\hat{H}-S)]|\Psi(\tau)\rangle.
\end{equation}
This equation can be projected onto the different determinants in the Hilbert space to give
\begin{equation}
    \langle D_{\bvec{i}}|\Psi(\tau+\delta\tau)\rangle=\langle D_{\bvec{i}}|\Psi(\tau)\rangle- \delta\tau \langle D_{\bvec{i}}|\hat{H}-S|\Psi(\tau)\rangle,
\end{equation}
which gives an update equation for the corresponding FCI parameters $c_{\bvec{i}}$, where $|\Psi(\tau)\rangle = \sum_{\bvec{i}}c_{\bvec{i}}(\tau)|D_{\bvec{i}}\rangle$:
\begin{equation}
    c_{\bvec{i}} (\tau + \delta\tau) = c_{\bvec{i}}(\tau) - \delta\tau[(H_{\bvec{ii}}-S)c_{\bvec{i}}(\tau) + \sum_{\bvec{j} \neq \bvec{i}}H_{\bvec{ij}} c_{\bvec{j}}(\tau)],
\end{equation}
where $H_{\bvec{ij}} = \langle D_{\bvec{i}}|\hat{H}|D_{\bvec{j}}\rangle$. This equation can be viewed as describing the population dynamics of particles placed on the different determinants and may be modelled by a stochastic process composed of three steps:\cite{booth_fermion_2009}
\begin{itemize}
    \item \textbf{Spawning:} Given determinant $D_{\bvec{i}}$, generate new particles on determinant $D_{\bvec{j}}$ with probability ${p \propto \delta\tau H_{\bvec{ij}} c_{\bvec{j}}(\tau)}$.
    \item \textbf{Death:} Given determinant $D_{\bvec{i}}$, generate new particles on determinant $D_{\bvec{i}}$ with probability ${p \propto \delta\tau (H_{\bvec{ii}} - S) c_{\bvec{i}}(\tau)}$
    \item \textbf{Annihilation:} For a given determinant, cancel out particles carrying opposite signs.
\end{itemize}
The formulation of CCMC closely matches that of FCIQMC, with the difference that instead of residing on determinants, walkers reside on \textit{excitors}, $\hat{a}_{\bvec{n}}$, defined as $\hat{a}_{\bvec{n}}|D_{\bvec{0}}\rangle= \pm|D_{\bvec{n}}\rangle $, {where the choice of sign is a matter of convention}. Replacing the FCI wavefunction by the coupled cluster \textit{ansatz} in \Cref{eq:linear_proj} and left-multiplying by $\langle D_{\bvec{i}}|$ gives
  \begin{equation}
  \label{eq:ccmc_master}
  \begin{split}
  \langle D_{\bvec{i}}|\Psi^{\mathrm{CC}}(\tau+\delta\tau) \rangle =& \langle D_{\bvec{i}}| \Psi^{\mathrm{CC}}(\tau)\rangle\\
  -&\delta\tau \langle D_{\bvec{i}}|(\hat{H}-S)|\Psi^{\mathrm{CC}}(\tau) \rangle,
  \end{split}
  \end{equation}

   The coupled cluster \textit{ansatz} parametrises the wavefunction with cluster amplitudes in a non-linear fashion. The mapping of CI coefficients to cluster amplitudes can be done by a simple projection, which reveals contributions from multiple clusters. For example in a CCSD wavefunction (\textit{i.e.}, ${\hat{T}}={\hat{T}}_1+{\hat{T}}_2 $)
  \begin{equation}
    \langle D_{ij}^{ab}|e^{{\hat{T}}}D_{\bvec{0}}\rangle = t_{ij}^{ab}+t_i^at_j^b-t_i^bt_j^a,
  \end{equation}
  with the negative sign arising from the fact that ${\hat{a}_b^{\dagger}\hat{a}_i\hat{a}_a^{\dagger}\hat{a}_j=-\hat{a}_a^{\dagger}\hat{a}_i\hat{a}_b^{\dagger}\hat{a}_j}$, due to the anti-commutation relations of the second-quantised creation and annihilation operators \cite{helgaker2014molecular}. Terms like $t_{ij}^{ab}$ are known as \textit{non-composite} cluster amplitudes, and the rest as \textit{composite} cluster amplitudes. Here we make the approximation that composite clusters have much smaller contributions than non-composite ones, their changes will be negligible per time step, and hence we can remove the $\mathcal{O}(\hat{T}^2) $ contributions on both sides to write
  \begin{equation}
  \label{eq:ccmc-master}
    t_{\bvec{i}}(\tau+\delta\tau) \rangle = t_{\bvec{i}}(\tau) - \delta\tau \langle D_{\bvec{i}}|(\hat{H}-S)|\Psi^{\mathrm{CC}}(\tau) \rangle.
  \end{equation}
  Compared to FCIQMC, an additional step needs to be performed for each Monte Carlo iteration: the sampling of the exponential \textit{ansatz}. For $N_{\mathrm{ex}}$ total walkers, also called \textit{excips} in CCMC, $\mathcal{O}(N_{\mathrm{ex}})$ clusters are formed randomly by combining present excitors according to specific biasing rules \cite{spencer_2016}.

  Finally, the intermediate normalisation \cite{mbpt_bartlett} of the wavefunction is redefined to give the CCMC \textit{ansatz}:
  \begin{equation}
  \label{eq:ccmc-ansztz}
    |\Psi_{\text{CCMC}}\rgl = N_0 e^{{\hat{T}}/N_0} |D_{\bvec{0}}\rangle,
  \end{equation}
  which introduces the reference population as a new independent variable, solving the problem that \Cref{eq:ccmc_master} does not lead to a viable update equation for $\langle D_{\bvec{i}}| = \langle D_{\bvec{0}}| $.

A multi-reference formulation of the CCMC algorithm (MR-CCMC) \cite{filip_multireference_2019} has been implemented, retaining a single-reference formalism, in common with the so-called  SRMRCC methods in Ref. \citenum{lyakh_multireference_2012}. The flexibility of the CCMC algorithm allows this multireference approach to be implemented with minimal code changes to the single-reference algorithm, bypassing what would be a challenging process in deterministic methods. Essentially, for a coupled cluster truncation level $P$, the algorithm allows any number of determinants to become a `secondary reference', stores excitors that are within $P$ excitations from \textit{any} references (instead of just the HF determinant), and allows clusters to form that are within $P+2$ excitations from \textit{any} reference. The set of references is commonly known as the model or reference space. To summarise, the algorithmic modifications relative to single-reference CCMC are:
\begin{itemize}
    \item Store all the secondary references in some searchable data structure, and additionally store the highest excitation level from the reference determinant among the secondary references, $k_{\text{max}}$.
    \item \textbf{Cluster expansion}: allow clusters with excitation level of up to $k_{\text{max}}+P+2$ to form, instead of $P+2$ in the single-reference case. Discard those that are not $P+2$ excitations away from \textit{some} reference determinant.
    \item \textbf{Spawning}: for a randomly generated spawnee (\textit{i.e.}, $\langle D_{\bm{j}}|$), check that it is within $P$ excitations of \textit{any} secondary references.
    \item \textbf{Cloning/death}: allow death on excitors that are within $P$ excitations from \textit{any} secondary references.
\end{itemize}

 While this MR-CCMC method can treat systems which conventional single-reference CC struggles with, this comes at an increased computational cost. Comparing contributor excitation levels to all references becomes expensive as the number of references grows, particularly when the contributor turns out to lie outside of the desired space. Therefore, non-trivial computational effort is expended on attempts that will not contribute to the overall estimators and propagation, while also making successful steps more expensive than their single-reference equivalents.
 
 In the rest of this paper, we will first introduce the wall-Chebyshev projector, {which can replace the traditional linear QMC projector,} and show that it can be applied to (MR-)CCMC and FCIQMC to reduce the number of times the Hamiltonian needs to be applied to reach statistical convergence, thereby reducing the computational cost. {\color{black}MR-CCMC in particular is a convenient testing ground for this new approach, as it can treat systems in a variety of correlation regimes, preserving polynomial scaling with system size, which makes calculations significantly cheaper than their FCIQMC counterparts. However, in order for the MR-CCMC algorithm to be able to take full advantage of the speed-up provided by the wall-Chebyshev projector, we also introduce a suite of specific modifications} to the MR-CCMC algorithm that accelerates the handling of the reference space. We apply the resulting algorithm to several {\color{black}traditional benchmark systems to investigate the performance enhancements due to the proposed algorithmic improvements.}

\section{The wall-Chebyshev projector}
\subsection{Motivation and theory}
In common projector-based QMC methods, including FCIQMC and CCMC, a linear projector (\Cref{eq:linear_proj}) is used. The first-order Taylor expansion turns out to be a very reasonable approximation, since we demonstrate in \Cref{appendix:no_benefit} that there is no benefit whatsoever in going to higher orders of the Taylor expansion of the exponential projector. However, this does not mean that one cannot devise more efficient projectors. An example is a projector based on a Chebyshev expansion of the wall function, which was first proposed in Ref. \citenum{zhang_2016} in the context of a deterministic projector-based selected CI algorithm. 

The wall function is given by
\begin{equation}
    {\text{wall}}(x)=
    \begin{cases}
        \infty,\ x< 0\\
        1,\ x=0\\
        0,\ x>0
    \end{cases},
\end{equation}
and is physically motivated as the infinite time limit of the exponential projector: 
\begin{equation}
    \lim_{\tau\rightarrow\infty}e^{-\tau (x-S)}=\text{wall}(x-S),
\end{equation}
which can map any trial wave function $|\Phi_0\rangle$ to the exact ground state $|\Psi_0\rangle$, if $\langle\Phi_0|\Psi_0\rangle\neq0$ and $E_0 \leq S < E_1$.

While a Taylor series expansion does not exist for the discontinuous wall function, an expansion in Chebyshev polynomials, like a Fourier expansion, is trivial. The Chebyshev polynomials of the first kind, defined as $T_n(\cos(\theta))=\cos(n\theta)$, form an orthogonal basis (with metric $(1-x^2)^{-1/2}$) for functions defined over $x\in[-1,1]$:
\begin{equation}
    \int^1_{-1}T_m(x)T_n(x)(1-x^2)^{-1/2}\dl x=\delta_{mn}\pi/(2-\delta_{m0}).
\end{equation}

To facilitate the following discussion, we define the \textit{spectral range}, $R$, of a Hamiltonian as $R=E_{N-1}-E_0$, where $E_i$ is the $i$-th eigenvalue of the full Hamiltonian and $N$ is the size of the Hilbert space. Furthermore, our energy range $\epsilon\in[E_0,E_{N-1}]$ requires the application of an affine transformation to the Chebyshev polynomials
\begin{equation}
    \epsilon = E_0+\frac{R}{2}(x+1),\ x\in[-1,1].
\end{equation}

We show in \Cref{appendix:wall-ch} that the $m$-th order Chebyshev expansion of the wall function is
\begin{equation}
\label{eq:wall-ch}
    g^{\text{wall-Ch}}_m(\epsilon) =\frac{1}{1+2m}\sum_{k=0}^m(2-\delta_{k0})T_k(-x).
\end{equation}

\begin{figure}[!htp]
    \centering
    \includegraphics[width=\columnwidth]{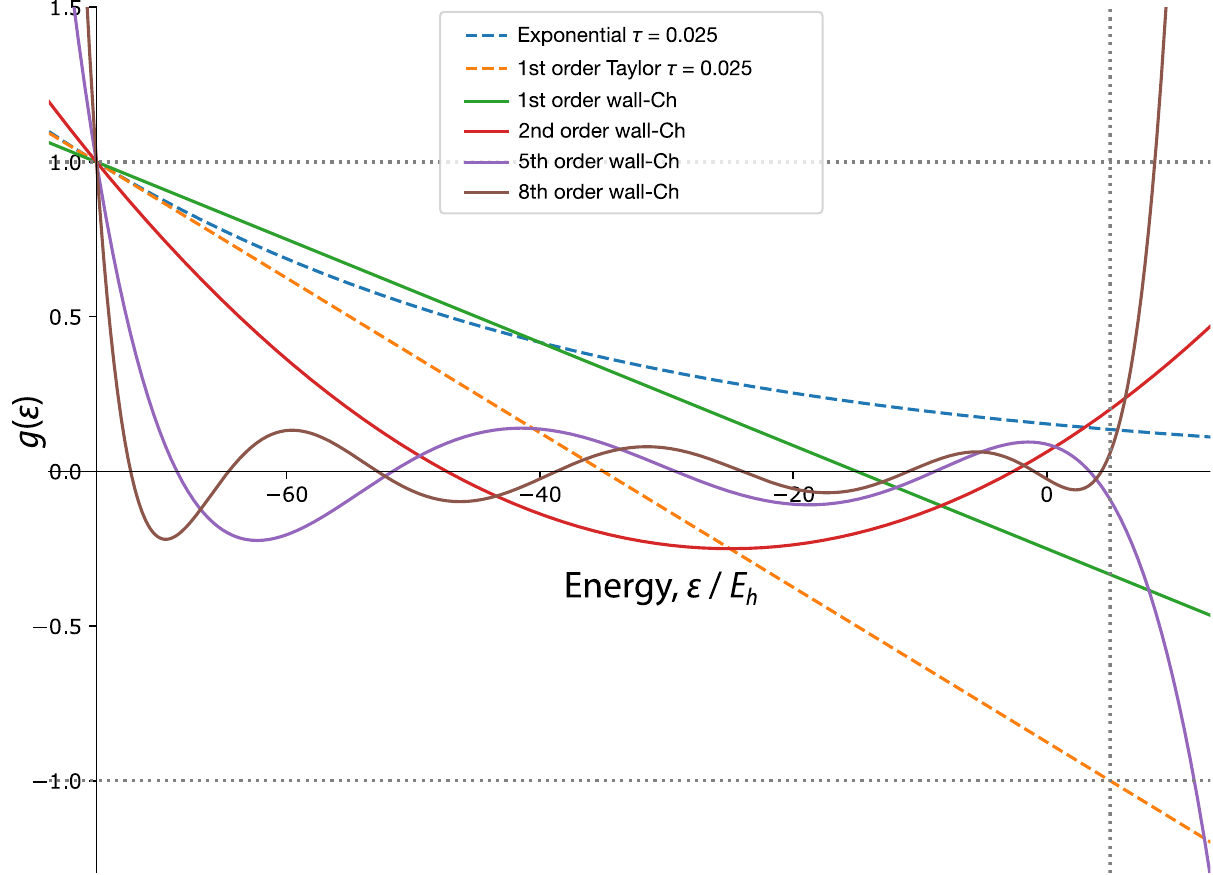}
    \caption{The Chebyshev expansions of the wall function in an arbitrary range of $[-75,5]$, compared to the linear projector with the maximal time step of $\delta\tau=0.025$, and its corresponding exponential projector.}
    \label{fig:wall-ch}
\end{figure}

For illustration purposes, we plot several orders of Chebyshev expansion in \Cref{fig:wall-ch}, where we can also observe the monotonic divergence to $+\infty$ for $\epsilon<E_0$. The other tail also diverges to $\pm\infty$ depending on the parity of the order $m$. 

In this instance, the nodes of \Cref{eq:wall-ch} are analytically known (see derivation in \Cref{appendix:wall-ch}) as 
\begin{equation}
    a_{\nu}=E_0+\frac{R}{2}\left(1-\cos\frac{\nu\pi}{m+1/2} \right).
\end{equation}
This allows us to decompose the $m$-th order projector into a product of $m$ linear projectors, each with their own weight that ensures $g^{\text{wall-Ch}}_m(E_0)=1$:
\begin{equation}
    g^{\text{wall-Ch}}_m(\epsilon) = \prod_{\nu=0}^{m-1} \frac{\epsilon-a_{\nu}}{E_0-a_{\nu}}.
\end{equation}

\begin{figure}[!htp]
    \centering
    \includegraphics[width=\columnwidth]{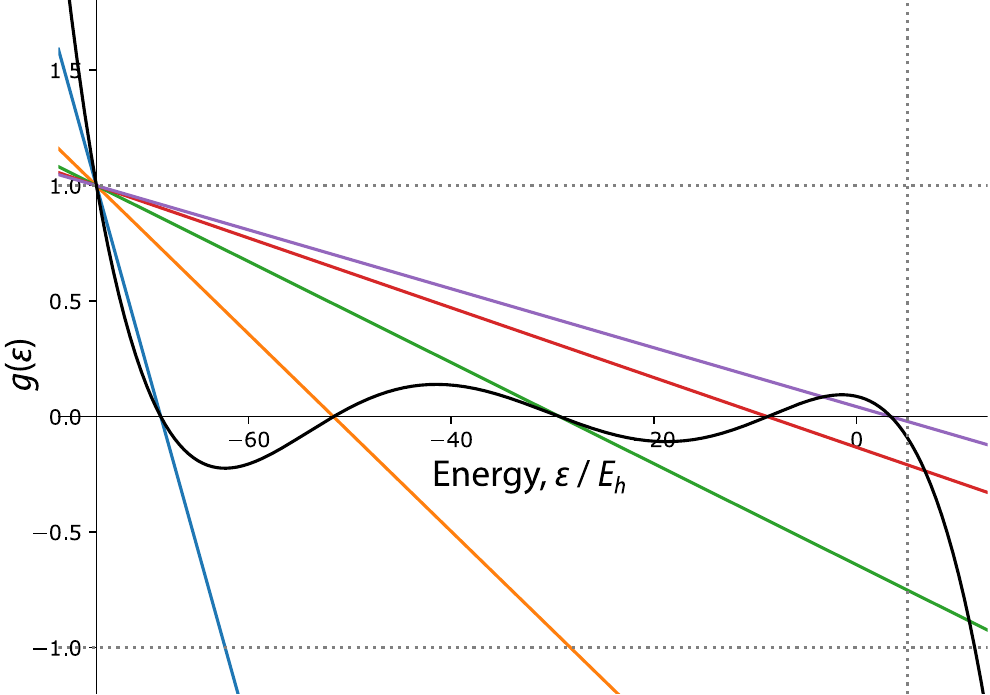}
    \caption{The fifth order Chebyshev expansion of the wall function, shown here to decompose into a product of $5$ linear projectors, each with their own effective time steps.}
    \label{fig:lin_prod}
\end{figure}

A decomposition for a fifth order Chebyshev expansion of the wall function can be seen in \Cref{fig:lin_prod}.

\subsection{Application to FCIQMC and CCMC}
\label{sec:cheby-ccmc}
In FCIQMC and CCMC, the lowest eigenvalue estimate is the shift, $S$, and the upper spectral bound can be a constant, estimated from the Gershgorin circle theorem \cite{gersgorin_uber_1931} as 
\begin{equation}
    \widetilde{E}_{N-1}=H_{N-1,N-1}+{{{\sum}^{\prime}_{{\bvec{j}}\in\{\mathbf{S},\mathbf{D}\}}}}H_{N-1,{\bvec{j}}},
    \label{eq:gersh}
\end{equation}
where the sum is over all determinants connected to the highest determinant (singles and doubles), and the `$'$'  restricts it to ${\bvec{j}\neq N-1}$. 

The action of the wall-Chebyshev projector on ${|\Psi^{(n,0)}\rangle=[g^{\text{wall-Ch}}(\hat{H})]^n|\Phi\rangle}$ is 
\begin{equation}
   |\Psi^{(n+1,0)}\rangle = g^{\text{wall-Ch}}(\hat{H})|\Psi^{(n,0)}\rangle,
\end{equation}
which gives the wavefunction after $n+1$ applications of the projector. We can additionally define the `intermediate' wavefunctions as
\begin{equation}
\label{eq:cheby_interm}
|\Psi^{(n,\mu)}\rangle = \left[\prod_{\nu=0}^{\mu}\frac{\hat{H}-a_{\nu}}{S-a_{\nu}}\right] |\Psi^{(n,0)}\rangle.
\end{equation}

We are now ready to derive the update equations for FCIQMC and CCMC. We start with the slightly more involved derivation for CCMC . Projecting these intermediate wavefunctions onto determinants, we have 
\begin{equation}
    \langle D_{\bvec{i}}|\Psi^{(n,\nu+1)} \rangle = \frac{1}{S-a_{\nu}} \langle D_{\bvec{i}}|{\hat{H}-a_{\nu}}|\Psi^{(n,\nu)} \rangle.
    \label{eq:shift_update}
\end{equation}
It is important now to distinguish between $\tilde{t}_{\bvec{i}}$, the projection of a wavefunction onto determinant $D_{\bvec{i}}$, and the corresponding excitor amplitude, $t_{\bvec{i}} $, with the former including unconnected (`composite') contributions. At convergence,

\begin{align}
\begin{split}
\tilde{t}_{\bvec{i}} = -\frac{1}{a_{\nu}-S} \left[\sum_{{\bvec{j}} \neq {\bvec{i}}} H_{\bvec{ij}} \tilde{t}_{\bvec{j}} + (H_{\bvec{ii}} - a_{\nu})\tilde{t}_{\bvec{i}}\right]&\\
\tilde{t}_{\bvec{i}} -t_{\bvec{i}}+t_{\bvec{i}} = -\frac{1}{a_{\nu}-S} \left[\sum_{{\bvec{j}} \neq {\bvec{i}}} H_{\bvec{ij}} \tilde{t}_{\bvec{j}} + (H_{\bvec{ii}} - a_{\nu})\tilde{t}_{\bvec{i}}\right]&\\
t_{\bvec{i}} = t_{\bvec{i}} - \frac{1}{a_{\nu}-S}\left[\sum_{\bvec{j}\neq\bvec{i}}H_{\bvec{ij}}\tilde{t}_{\bvec{j}} +(H_{\bvec{ii}}-S)\tilde{t}_{\bvec{i}} \right]&
\end{split}
\end{align}
We may now convert the last equation into an update step,
\begin{equation}
    t_{\bvec{i}}(\tau + \delta\tau) = t_{\bvec{i}}(\tau) - \frac{1}{a_{\nu}-S}\left[\sum_{\bvec{j}\neq\bvec{i}}H_{\bvec{ij}}\tilde{t}_{\bvec{j}}(\tau) +(H_{\bvec{ii}}-S)\tilde{t}_{\bvec{i}}(\tau) \right]
\end{equation}
Comparing with the original update equations, which are given by
\begin{equation}
    t_{\bvec{i}}(\tau + \delta\tau) = t_{\bvec{i}}(\tau) - \delta\tau\left[\sum_{\bvec{j}\neq\bvec{i}}H_{\bvec{ij}}\tilde{t}_{\bvec{j}}(\tau) +(H_{\bvec{jj}}-S)\tilde{t}_{\bvec{i}}(\tau) \right],
\end{equation}
we reach the conclusion that the necessary modifications are
\begin{enumerate}
    \item Setting $\delta\tau=1$
    \item Applying the $m$ constituent linear projectors in the $m$-th order wall-Chebyshev projector. For linear projector ${\nu\in \{0,\dots,m-1\}}$, scale Hamiltonian elements in spawning and death by $1/(a_{\nu}-S)$ (`Chebyshev weights').
\end{enumerate}
The same analysis can be performed on FCIQMC, without the complication of composite amplitudes, to obtain a similar set of update equations:
\begin{equation}
    c_{\bvec{i}} = c_{\bvec{i}}-\frac{1}{a_{\nu}-S}\left[\sum_{\bvec{j}\neq\bvec{i}}H_{\bvec{ij}}c_{\bvec{j}}+(H_{\bvec{ii}}-S)c_{\bvec{i}} \right].
\end{equation}
In terms of implementation, the two sets of update equations are nearly identical, and can share the same code in large parts.

Analysis of the asymptotic rate of convergence (see \Cref{appendix:conv}) shows that the theoretical speedup of an order $m$ wall-Chebyshev projector relative to the linear projector with largest allowed $\delta\tau$ is $(m+1)/3$. \cite{zhang_2016} Due to blooms, the largest $\delta\tau$ is never reached in the conventional propagator, so real speedups are expected to be larger.

\subsection{The shift update procedure}
The original shift update equation for CCMC and FCIQMC is given by
\begin{equation}
    S^{(n+A)}=S^{(n)}-\frac{\zeta}{A\delta\tau}\ln\left( \frac{N_{\mathrm{w}}^{(n+A)}}{N_{\mathrm{w}}^{(n)}} \right),
\end{equation}
where the update is performed every $A$ time steps, $\zeta$ is the shift damping parameter, and $N_{\mathrm{w}}$ is the total number of walkers.

Due to the time step $\delta\tau$ being set to unity, the shift update procedure is expected to become rather unresponsive to the changes in particle population. As a consequence, there can be vastly uncontained spawning, unchecked by the lower-than-expected deaths, resulting in unmanageable population growths. To remedy this, initially, a scaled update procedure was experimented with, setting $A=1$:
\begin{equation}
\label{eq:wrong_pop_control}
    S^{(n+1)}=S^{(n)} -{\zeta}\sum_{\nu=0}^{\mu}[a_{\nu}^{(n)}-S^{(n)}]\ln\frac{N_{\mathrm{w}}^{(n,\nu)}}{N_{\mathrm{w}}^{(n,\nu-1)}},
\end{equation}
which seemed attractive as it reduces to \Cref{eq:shift_update} in the first-order case where the sum only contains one term or if all Chebyshev weights are the same. However, this was not successful in reining in the population growth. We believe this is because the intermediate wavefunctions in \Cref{eq:cheby_interm} are ill-behaved due to being generated by an effective time step potentially larger than $\tau_{\mathrm{max}}$. A series of population changes that start and end at $N^{(n)}$ and $N^{(n+1)}$ respectively can produce very different values of shift update in \Cref{eq:wrong_pop_control}, and the shift update produced is very sensitive to the unreliable intermediate values. Hence the population information from these intermediate wavefunctions should not be used.  
\iffalse
\begin{figure}[!htp]
    \centering
    \includegraphics[width=\columnwidth]{cheby.unresponsive.png}
    \caption{\ce{C2}/cc-pVDZ at $2.0$ \AA~separation, with a fifth-order wall-Chebyshev projector. With default settings, the Chebyshev propagator can cause the shift to become unresponsive to the large changes in total population, resulting in uncontained population growth and slow convergence. \color{black} Can we add the total population here?}
    \label{fig:cheby_unresponsive}
\end{figure}
\fi

Another procedure that was more successful was to decrease the damping (by increasing $\zeta$) of the shift updates, causing the shift to be more responsive to the changes in populations, which in turn helps stabilise the population. We also found it helpful to use the improved shift update procedure outlined in Ref. \citenum{yang_improved_2020}, where an additional term is added to the shift-update procedure:
\begin{equation}
    S^{(n+A)} \leftarrow -\frac{\xi}{A\delta\tau}\ln\left(\frac{N_{\mathrm{w}}^{(n+A)}}{N_t} \right),
\end{equation}
where $\xi$ is the `forcing strength', and $N_t$ is the target population. This has the effect of additionally stabilising the population by `pinning' it to the pre-set target population. A further proposal from the same paper, arising from an argument from a scalar model of population dynamics, is for critical damping to be achieved by setting $\xi=\zeta^2/4$. This is also found to be helpful. Altogether, these modifications result in greatly improved population control and we were able to obtain dynamics that can be used in a reblocking analysis, as shown in \Cref{fig:c2-cheby-taumax-fair}, for example. 

In practice, we have also found that with increasing order of Chebyshev projector, a larger target population is usually needed, otherwise the calculation may exhibit a sign-problem-like divergence. This may be attributed to the larger effective time steps that the higher order projectors use and is documented elsewhere, for example, see Fig. 2 in Ref. \citenum{vigor_understanding_2016}. 

\section{Accelerating the MR-CCMC algorithm}
\label{sec:acc}
{\color{black}The MR-CCMC method is a promising candidate for tackling strongly correlated systems at polynomial cost, and represents an economical alternative to the related exponentially scaling FCIQMC method. In this section we detail two algorithmic developments that have greatly accelerated the MR-CCMC calculations in the remainder of the article, and have brought MR-CCMC a step closer to algorithmic maturation.}
\subsection{Efficient cluster acceptance algorithm}
In the spawning step of the MR-CCMC algorithm, we check that a spawnee is within $P$ excitations of \textit{any} secondary reference. The same check needs to be performed in the death step. The original MR-CCMC algorithm performed a linear scan through the list of secondary references, which is clearly a $\mathcal{O}(n_{\mathrm{ref}})$ operation, where $n_{\mathrm{ref}}$ is the number of secondary references. The subroutine that decides whether a spawn is accepted is the second most frequently called subroutine in the program, after the excitation generator. Therefore, a linear search in this step can quickly become prohibitively expensive in a moderate to large reference space ($n_{\mathrm{ref}}>1000$, for example). We note that traditional data structures and search algorithms, such as a binary search on a sorted list of secondary references, would not work here, as the definition of `distance' in this case, \textit{i.e.}, excitation-rank, is non-Euclidean. A search algorithm in a general metric space is therefore needed. 

\begin{figure}[!htp]
    \centering
    \includegraphics[width=0.8\columnwidth]{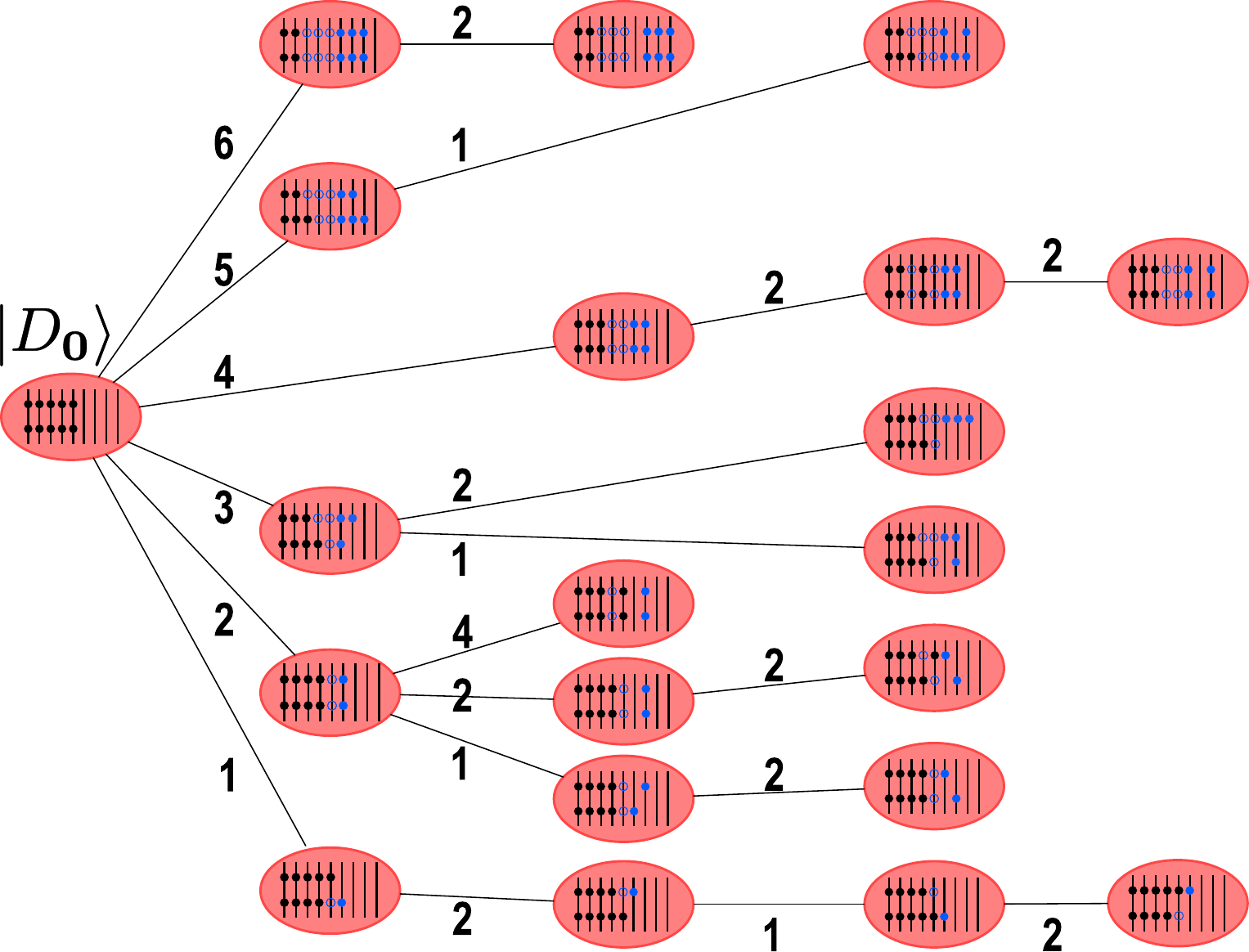}
    \caption{The BK tree can conduct efficient nearest neighbour searches in a discrete metric space, like the excitation rank. In this figure a BK tree built from $20$ arbitrary determinants is shown. The topology of the tree is not unique, and is dependent on the order the nodes were added to the tree.}
    \label{fig:bk_tree}
\end{figure}

\begin{figure*}[!htp]
    \centering
    \includegraphics[width=0.8\textwidth]{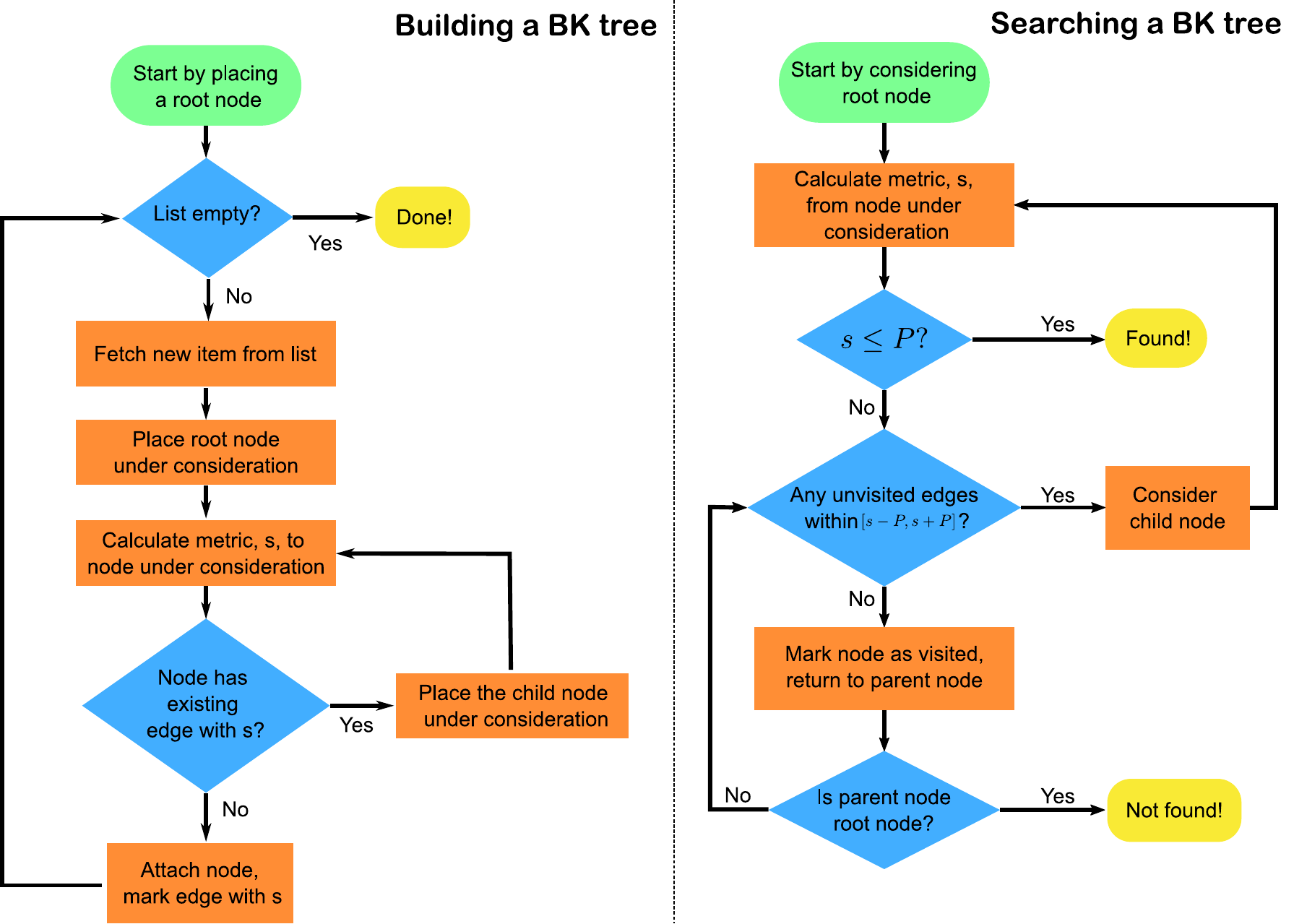}
    \caption{Flowcharts for the building and searching of a BK-tree.}
    \label{fig:bk_tree_algo}
\end{figure*}

The excitation-rank distance between two Slater determinants is equivalent to half the Hamming distance between the bit strings representing these two determinants, and the Hamming distance is a well-known example of a discrete metric \cite{hamming_error_1950}. A data structure, known as the BK tree \cite{burkhard_approaches_1973}, is particularly well suited for efficient searches in discrete metric spaces. The tree, an example of which is given in \Cref{fig:bk_tree}, is constructed only once at the beginning of the calculation. Subsequently, the search can be performed in $\mathcal{O}(\log n_{\mathrm{ref}})$ time using a recursive tree traversal algorithm. The tree construction and search algorithms are pictured in \Cref{fig:bk_tree_algo}.

\subsection{Compression of the reference space}
\label{sec:bk-comp}
Whereas many classical multireference coupled cluster (MRCC) methods work with complete active spaces (CAS), the MR-CCMC algorithm as presented here is highly flexible as to the shape of the reference space, and as such can be considered a general reference space (GMS) method \cite{lyakh_multireference_2012}. This enables us to consider arbitrary subsets of the CAS as the reference space, and fine-tune the balance between cost and accuracy. One of us has devised a compression method in this vein \cite{zhaoDevelopmentsLargeScale2022}. Here we briefly summarise its main thrust.

Two of us observed that \cite{filip_multireference_2019} that for some $(ne, no)$ active space, the results of a MR-CCSD calculation using all of the determinants in this active space as references (\textit{i.e.}, a CAS MR-CCSD calculation) is qualitatively very similar to the results of a MR-CCSD\dots$m$ calculation, where $m=n/2$, using only the `bottom' and `top' determinants (\textit{i.e.}, the \textit{aufbau} and anti-\textit{aufbau} determinant respectively) of the CAS as the references. We term the latter calculation as `2r-CCSD\dots$m$'. Using this observation, we aim to algorithmically generate \textit{only} those determinants that are in the Hilbert spaces of both the CAS MR-CCSD and 2r-CCSD\dots$m$ calculations, which should provide us with a compressed set of reference determinants that captures the most important determinants in the CAS. It was shown that this set of determinants can be generated by enumerating determinants of up to $(m-2)$--fold excitations from the bottom and top determinants.

\section{Computational details}

\label{sec:pg_symmetry}
In this work we study the carbon and beryllium dimers, {\color{black} using MR-CCMC and initiator FCIQMC (i-FCIQMC).\cite{cleland_2010} The first system displays challenging multireference behaviour, requiring an ($8e,8o$) CAS as the reference space for MR-CCMC, which is large enough to benefit from the techniques presented in \Cref{sec:acc}. Overall MR-CCMC and i-FCIQMC calculations are performed in the full space of ($12e, 28o$). The beryllium dimer on the other hand is only moderately multireference, but exhibits weak bonding, with a dissociation energy of only approximately 4 m$E_{\mathrm{h}}$. Accurately describing this behaviour in QMC requires low stochastic noise in the energy estimates. The accelerated convergence provided by the wall-Chebyshev propagator is therefore particularly beneficial in reducing the propagation time required to obtain low-variance estimates.} For these systems, Dunning's cc-pVXZ basis sets are used \cite{dunning_gaussian_1989}. The required electron integrals are generated by the Psi4 \cite{smith_psi4_2020} and \textsc{PySCF} \cite{sun_pyscf_2018} packages. The electron integrals are generated in the $D_{2h}$ point group symmetry and transformed into the basis of $\hat{L}_z$ eigenfunctions based on the \texttt{TransLz.f90} script provided in the \textsc{NECI} package \cite{guther_neci_2020}, which we re-wrote to interface with \textsc{PySCF}. The `heat bath' excitation generator \cite{holmes_efficient_2016} is used whenever possible, otherwise the renormalised excitation generator \cite{booth_linear-scaling_2014} is used. 

The use of the $\hat{L}_z$ eigenfunctions helps not only further reduce the size of the relevant symmetry sector, but also helps distinguish low-lying states that would descend to the same irreducible representation in $D_{2h}$. For \ce{C2}, this would be the $^1\Sigma_g^+$ state and the $^1\Delta_g$ states, which both descend into the $^1A_g$ state in $D_{2h}$. The two states approach and cross each other \cite{wouters_chemps2_2014,sharma_general_2015}, which would prove challenging, if not impossible, to distinguish in $D_{2h}$.

When employing the wall-Chebyshev projector, the upper spectral range estimate obtained from the Gershgorin theorem (\Cref{eq:gersh}) is scaled up by $10\%$ by default to guarantee an upper bound on the spectral width of the Hamiltonian. {\color{black}For i-FCIQMC applications, we found increasing this factor to $50\%$ improved population dynamics.}

Quantum Monte Carlo calculations are carried out using the \textsc{HANDE-QMC} package\cite{spencer_hande-qmc_2019}.

\section{Results and discussion}
\label{sec:bk_results}
\subsection{Reference space treatment in MR-CCMC}
\subsubsection{BK-tree search}
For a \ce{C2} system with a full $(8e,8o)$ CAS as the reference space without symmetry screening ($4900$ references that preserve the $M_s=0$ symmetry), the BK-tree search is benchmarked against a na\"ive linear search, which loops over all secondary references and terminates when one of the references is within $P$ excitations of the target determinant. The validity of the BK-tree search is separately established by performing a normal calculation with either search algorithm using the same random number generator seed, and asserting that the results are the same. Benchmarking results are given in \Cref{tab:search_comp}.
\begin{table}[!htp]
\centering
\caption{Timing comparison between the BK tree and naive search algorithms, for \ce{C2} using a $(8e,8o)$ CAS as the reference space for a multireference CCMCSD calculation.}
\begin{tabular}{lcc}
\hline
        & Overall timing / s & Time per spawning attempt / $\mu$s \\ \hline
BK tree & $809.28$           & $12.761$                           \\
Linear  & $5995.16$          & $94.533$                           \\ \hline
\end{tabular}

\label{tab:search_comp}
\end{table}

An apparent $8\times$ speedup is observed. Without performing a full profiling study, the actual reduction in time cost of the acceptance search is expected to be greater than $8\times$ as a complex series of operations is performed per spawning attempt on top of the acceptance search.

\subsubsection{Reference space compression}

\begin{figure}[!htp]
    \centering
    \includegraphics[width=\columnwidth]{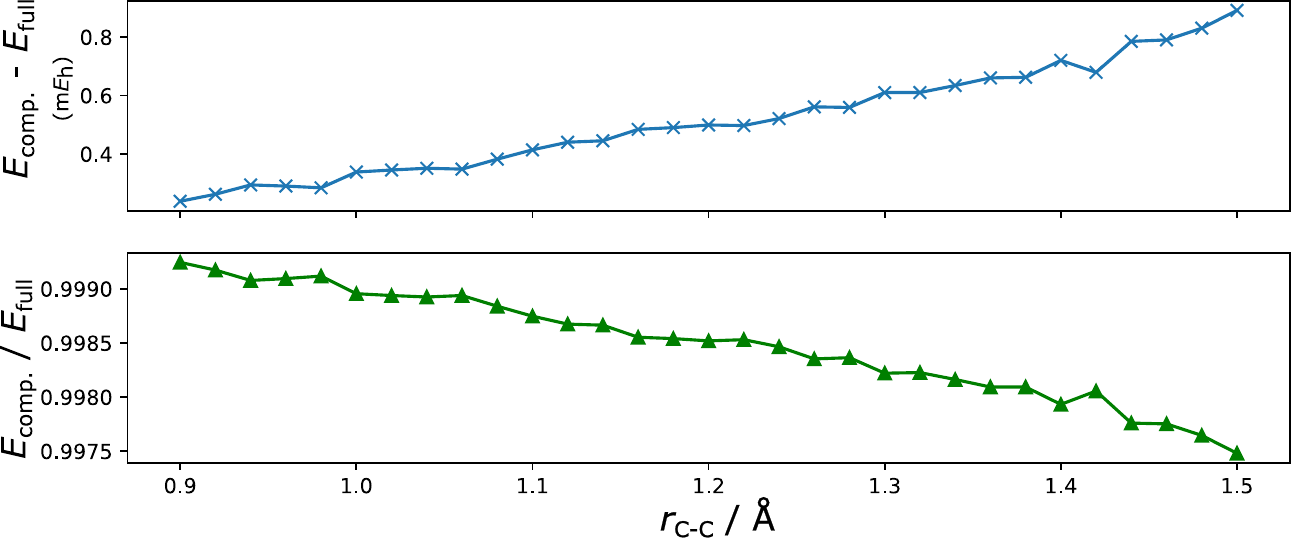}
    \caption{The correlation energy for \ce{C2}/cc-pVDZ at $r_\mathrm{C-C} = 0.9-1.5$\AA~with the compressed set of $721$ secondary references, relative to {using} the full CAS reference space. We observe that, despite a $7$-fold reduction in the size of the reference space, the reductions in the correlation energy captured are much smaller, making this an attractive trade-off. The stochastic error bars are too small to be seen, due to the use of the semi-stochastic algorithm \cite{zhaoDevelopmentsLargeScale2022}.}
    \label{fig:compress}
\end{figure}

For the $(8e,8o)$ CAS used for \ce{C2}, the compression method discussed in \Cref{sec:bk-comp} yields a total of $722$ determinants in the compressed reference space. Here we show the results for the \ce{C2}/cc-pVDZ system at separations of $0.9$ to $1.5$ \AA. The performance of the compression scheme is shown in \Cref{fig:compress}. Here we have employed the default quasi-Newton acceleration\cite{neufeld_2020} implemented in \textsc{HANDE}. We can see that despite the almost $7$-fold reduction in the reference space (and consequently a similar reduction in computational cost), the errors are within chemical accuracy ($1.6$ m$E_h$).

\subsubsection{Binding curve of the carbon dimer}
\begin{figure}[!htp]
    \centering
    \includegraphics[width=\columnwidth]{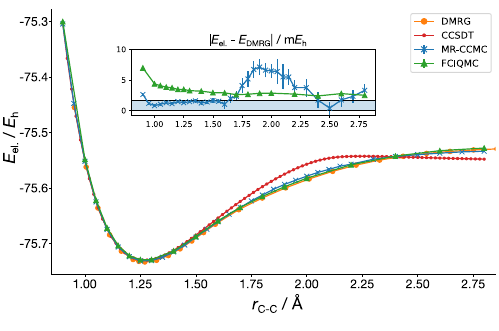}
    \caption{The binding curve of the $^1\Sigma_g^+$ state of \ce{C2}/cc-pVDZ in the range of $0.9$-$2.8$ \AA~separation. {All-electron} MR-CCMCSD calculations are based on CASSCF orbitals and use an $(8e,8o)$ CAS as a reference space, with clusters truncated at the double excitation level from this. The FCIQMC data is from Ref. \citenum{booth_breaking_2011}, and the DMRG data is from \citenum{wouters_chemps2_2014}, and the CCSDT data is from the \textsc{ccpy} package developed by Piecuch and coworkers.\cite{ccpy} {\color{black}The inset shows the error in the MR-CCMC and FCIQMC energy relative to DMRG.}}
    \label{fig:c2-binding-curve}
\end{figure}

Finally, we studied the {\color{black}$\mathrm{X}\ ^1\Sigma_g^+$} state of the carbon dimer in the cc-pVDZ basis using MR-CCMCSD with these accelerations. The carbon dimer is a challenging test case for electronic structure methods, and the challenge is two-fold: firstly, as mentioned in \Cref{sec:pg_symmetry}, the {\color{black}$\mathrm{X}\ ^1\Sigma_g^+$} state becomes very nearly degenerate with the exceptionally low-lying {\color{black}$\mathrm{B}\ ^1\Delta_g$} state at stretched bond lengths,\cite{abramsFullConfigurationInteraction2004} and both states descend to the $A_g$ state in the commonly used $D_{2h}$ point group symmetry, rendering it very challenging to distinguish both states without the use of the $L_z$ symmetry, with one paper resorting to tracking individual CI coefficients \cite{wouters_chemps2_2014}; secondly, there is an abundance of avoided crossings, and specifically, the first excited {\color{black}${\mathrm{B}'}\ ^1\Sigma_g^+$} state participates\cite{varandas_extrapolation_2008} in an avoided crossing with the ground state at a bond length of around $1.6$ \AA~, resulting in a change in the most highly weighted diabatic state (\textit{i.e.}, determinant). This makes MR-CCMC calculations based on RHF orbitals exhibit long-imaginary-time instabilities for stretched geometries, which preclude obtaining accurate estimators. The binding curve presented in \Cref{fig:c2-binding-curve} used the full $(8e,8o)$ CAS as the {reference} space for a MR-CCMCSD calculation, with orbitals obtained using \textsc{PySCF} from an $(8e,8o)$ state-average CASSCF calculation over the lowest three $^1A_g$ states (in the $D_{2h}$ point group). The orbital coefficients are still tagged with their corresponding $D_{\infty h}$ irreducible representations, enabling us to perform the $L_z$ transformation. We ensured that the $\pi_u$ manifold was included in the reference determinant that generates the Hilbert space and secondary references.
Non-parallelity errors (NPE), defined here as the difference between the maximal and minimal deviation from the DMRG energies, are shown in \Cref{tab:npe}.

\begin{table}[!htp]
    \centering
    \begin{tabular}{llll}
    \hhline{====}
           & NPE / m$E_{\mathrm{h}}$ & Max AD /m$E_{\mathrm{h}}$ & Min AD /m$E_{\mathrm{h}}$\\\hline
    FCIQMC & 4.6 & 6.9 (0.9\AA) & 2.4 (2.4 \AA) \\
    MRCCMC   &10.4&7.1 (1.9\AA)&0.4 (2.5\AA)\\
    CCSDT  &45.5 &28.1 (2.0\AA) & 0.3 (2.42\AA)\\ \hhline{====}
    \end{tabular}
        \caption{The non-parallelity error, maximal, and minimal absolute deviations of the carbon dimer binding curve calculated with MR-CCMCSD using a $(8e, 8o)$ CAS reference space, {\color{black} FCIQMC and CCSDT} compared to the DMRG results. The numbers in parentheses indicate the bond length (in angstrom) at which the maximal/minimal absolute deviations occur.}
    \label{tab:npe}
\end{table}

\subsection{Chebyshev Propagator Results}

\begin{figure}[!htp]
    \centering
    \includegraphics[width=\columnwidth]{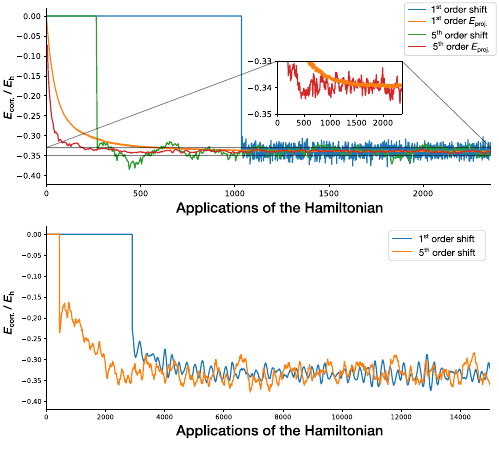}
    \caption{\color{black}QMC simulations of \ce{C2}/cc-pVDZ at $1.2$ \AA~separation, using the Chebyshev propagator. The top panel shows MR-CCMCSD calculations with a full $(8e,8o)$ CAS reference space, run with the first (simulated) and fifth order Chebyshev projectors. The fifth order calculation used a target population of $2\times10^6$ and a shift damping parameter of $0.5$, and the first order calculations used a target population of $1\times10^6$, and a shift damping parameter of $0.05$. The inset shows that the projected energy only barely stabilises around the true value at the end of the calculation using the linear projector. The bottom panel shows i-FCIQMC calculations, run with the first and fifth order Chebyshev projectors. Both calculations used target populations of $2\times10^6$ and a shift damping parameter of $0.5$. The projected energy estimator for high-order wall-Chebyshev FCIQMC displays higher noise than the shift, so we do not show it here for clarity. All calculations were carried out with a two-stage harmonic forcing shift update scheme. }
    \label{fig:c2-cheby-taumax-fair}
\end{figure}

\Cref{fig:c2-cheby-taumax-fair} shows an example of the power of the Chebyshev propagation, applied to {\color{black}MR-CCMCSD and i-FCIQMC calculations for \ce{C2}. In MR-CCMCSD, the shoulder height is reached after around $50$ iterations with the 5th order wall-Chebyshev propagator, even with a higher target population than the corresponding 1st order calculation. The dynamics is equilibriated essentially instantaneously,} which means all that is left to do is collecting statistics. On an Intel(R) Xeon(R) E5-2650 v2 CPU, the calculation in the figure was run for only $2$ hours on $6$ physical cores. Without the Chebyshev projector, the same calculation takes around $24$ hours with $12$ physical cores to give the same statistical error bar. 

{\color{black}The same trend can be observed in the i-FCIQMC calculation. We note that, due to the formally large time-step employed in wall-Chebyshev propagation, larger initiator thresholds are needed to easily stabilise calculations at low walker numbers than in conventional FCIQMC.}

\iffalse
\begin{figure}[!htp]
    \centering
    \includegraphics[width=\columnwidth]{c2-cheby.pdf}
    \caption{A fifth order Chebyshev projector applied to MR-CCMCSD for \ce{C2}/cc-pVDZ at $1.2$ \AA~separation, with a full $(8e,8o)$ CAS as the reference space, with a target population of $2\times10^6$ and a shift damping parameter of $0.5$, as compared to a linear projector with $\delta\tau=0.001$, a target population of $1\times10^6$, and a shift damping parameter of $0.05$. Both calculations are carried out with a two-stage harmonic forcing shift update scheme. The inset shows that the projected energy only barely stabilises around the true value at the end of the calculation using the linear projector. This means that the reblocking algorithm will disregard essentially all of the iterations shown here, due to the shift and the projected energy not agreeing.}
    \label{fig:cheby_c2}
\end{figure}
\fi

\begin{figure}[!htp]
    \centering
    \includegraphics[width=\columnwidth]{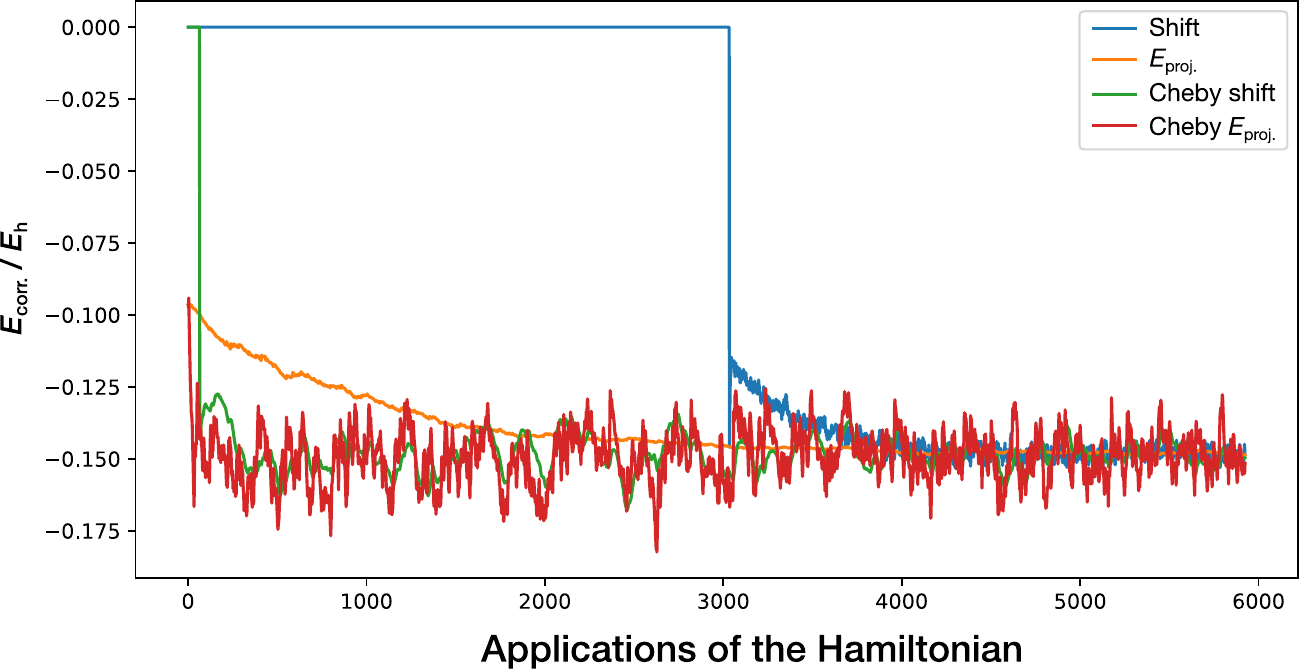}
    \caption{The second order projector (green and red lines) and the default linear projector (blue and orange lines) at $\delta\tau=0.002$ applied to MR-CCMCSD for the \ce{Be2}/cc-pVQZ system at $2.5$ \AA~separation, with a symmetry-screened $(4e,8o)$ (full $2s,2p$ valence) CAS as the reference space. Both calculations have a target population of $3\times10^6$.}
    \label{fig:be2-cheby}
\end{figure}

The following example shows \ce{Be2}, a smaller, modestly multi-reference system, {\color{black}which however requires the inclusion of contributions beyond doubles from the HF reference to get a qualitatively correct binding curve. In \Cref{fig:be2-cheby},} compared with the linear projector with a guessed $\delta\tau=0.002$, the second-order Chebyshev expansion shows a clear speed-up in convergence. In fact, the Chebyshev calculation took $66$ applications of the Hamiltonian to reach the target population of $3\times10^6$, whereas the linear projector took $3034$ applications to reach the same target population.

\begin{figure}[!htp]
    \centering
    \includegraphics[width=\columnwidth]{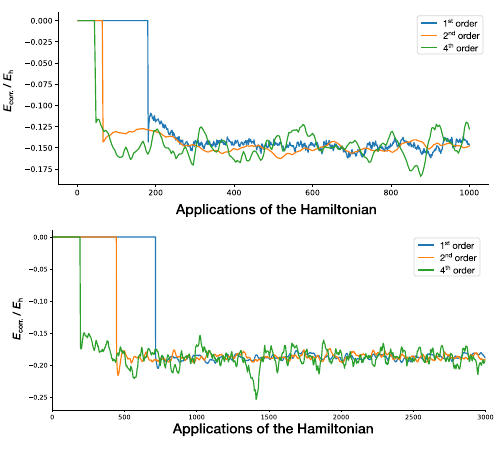}
    \caption{\color{black}The top panel shows an MR-CCMCSD propagation for \ce{Be2}/cc-pVQZ at $2.5$ \AA~separation with the $(4e,8o)$ CAS reference space run with the first, second and fourth order Chebyshev projectors. The bottom panel shows an i-FCIQMC propagation of \ce{Be2}/cc-pVTZ at $2.5$ \AA~separation, using the same propagators. Here we only show the shift as an estimator for $E_\mathrm{corr}$ for clarity.}
    \label{fig:be2-1-2-4-cheby}
\end{figure}

However, the second-order Chebyshev projector is expected to be as efficient as the linear projector with the maximum allowed time step (see \Cref{sec:cheby-ccmc}). To provide a fairer comparison, we present in \Cref{fig:be2-1-2-4-cheby}  the first, second and fourth order Chebyshev projector applied to {MR-CCMCSD and i-FCIQMC calculations using for the \ce{Be2} system at $2.5$ \AA, in the cc-pVQZ and cc-pVTZ basis sets, respectively.} The first order Chebyshev projector is equivalent to a linear projector with $\delta\tau=3/(E_{N-1}-E_0)$, which is $2/3$ of $\delta\tau_{\text{max}}$, but this maximal time step is commonly found to give rise to destabilising population blooms. Part of the benefit of our Chebyshev propagator algorithm is the automatic determination of the effective time steps, or `Chebyshev weights', \textit{via} the Gershgorin circle theorem, which in the first-order expansion limit reduces to an automatic way of choosing a good time step, instead of using trial and error.

In \Cref{fig:be2-1-2-4-cheby} we can clearly see the reduction in time needed to reach the shoulder height. It is worth bearing in mind that for MR-CCMCSD the three calculations require different target populations to stabilise, with the first and second order projectors having target populations of $3\times10^6$, and the fourth order projector having a target population of $5\times10^6$, which slightly increases the number of iterations required to reach the target population. {\color{black} In i-FCIQMC, despite using the smaller cc-pVTZ basis set, a target population of $5\times10^6$ is used for all calculations.}

\begin{figure}[!htp]
    \centering
    \includegraphics[width=\columnwidth]{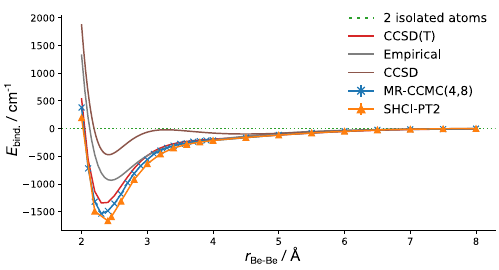}
    \caption{The binding curve of the beryllium dimer using the cc-pVTZ basis, computed using the CCSD, CCSD(T), MR-CCMCSD with a $(4e, 8o)$ CAS, and SHCI-PT2 methods. The empirical binding curve from \cite{sheng_analyzing_2013,patkowskiElusiveTwelfthVibrational2009} is also shown for comparison.}
    \label{fig:be2-tz}
\end{figure}

Finally, as a demonstration of the applicability of the Chebyshev projector in different correlation regimes, we have computed the binding curve of the \ce{Be2}/cc-pVTZ system using a $(4e,8o)$ CAS as a reference space for MR-CCMCSD, using the fifth-order Chebyshev projector. CCSD(T) and semistochastic heat-bath configuration interaction with second-order perturbation correction (SHCI-PT2) \cite{sharmaSemistochasticHeatBathConfiguration2017} results are also shown. The CCSD(T) results are from \textsc{Psi4},\cite{smith_psi4_2020}, and the SHCI-PT2 results are generated using the \textsc{Dice} plug-in to the \textsc{PySCF} package\cite{sun_pyscf_2018}. For SHCI-PT2, the full (8e, 28o) space was correlated, and hence it can be considered a surrogate for FCI results. The variational threshold was set to $\epsilon_1=8\times10^{-5}$ $E_{\mathrm{h}}$, and the PT2 threshold was set to $\epsilon_2=1\times10^{-8}$ $E_{\mathrm{h}}$, using $N_d=200$ deterministic determinants, with 5 PT2 iterations. \Cref{fig:be2-tz} shows the binding curves. The cc-pVTZ basis is known to severely overbind the beryllium dimer,\cite{guther_binding_2021} compared to the experimental value of $934.9 \pm 2.5$ cm$^{-1}$.\cite{merritt_beryllium_2009,meshkov_direct-potential-fit_2014,patkowskiElusiveTwelfthVibrational2009} Notwithstanding, the MR-CCMC method using the Chebyshev projector was able to provide consistently better description of the binding curve than the `gold-standard' CCSD(T). The MR-CCMC results are close to but not qualitatively the same as FCI results near the equilibrium, where static correlation dominates,\cite{el_khatib_beryllium_2014}, and are near-identical to FCI results in the stretched region, where dynamic correlation dominates, and a compact coupled-cluster representation of the wavefunction is beneficial.

\section{Conclusion}

We present here a series of algorithmic changes that can be used to accelerate the MR-CCMC algorithm in particular and QMC algorithms in general.

Specific to the MR-CCMC algorithm, we have introduced a BK-tree-based search algorithm to verify whether proposed clusters and spawns are within the accepted manifold for a given reference space. This reduces the scaling of this step from $\mathcal{O}(n_\mathrm{ref})$ to $\mathcal{O}(\log n_\mathrm{ref})$, which translates to an $8\times$ speed-up for the molecular systems studied. We have also designed a compression method for the reference space, which preserves only what we expect to be the most significant reference determinants. This decreases the size of the space by close to an order of magnitude. Finally, we have shown that only including references that belong to the same symmetry sector as the desired solution is also effective as a means to reduce the size of the reference space, while introducing only negligible additional error to the results.

We have also developed a new projector based on the Chebyshev polynomial expansion of the wall function, which significantly accelerates the convergence of QMC calculations. In an example calculation on the \ce{Be2} molecule, this reduced the number of Hamiltonian applications needed to reach the target population by two orders of magnitude. The wall-Chebyshev projector is generally applicable to different QMC algorithms, including FCIQMC and (MR-)CCMC approaches, so we believe that, together with many recent developments in increasing the apparent scarcity of the Hamiltonian and optimising the shift behaviour,\cite{Ghanem2019,Ghanem2020,yang_improved_2020}it is a promising step in expanding the range of applications for these methods. 

\section*{Author contributions}
%We strongly encourage authors to include author contributions and recommend using \href{https://casrai.org/credit/}{CRediT} for standardised contribution descriptions. Please refer to our general \href{https://www.rsc.org/journals-books-databases/journal-authors-reviewers/author-responsibilities/}{author guidelines} for more information about authorship.
ZZ, MAF and AJWT all contributed to the conceptualization of the project. ZZ developed the necessary software and analysed the data, under supervision from MAF and AJWT. The manuscript was written by ZZ and MAF, with input from all authors.

\section*{Conflicts of interest}
There are no conflicts to declare.

\section*{Acknowledgements}
ZZ was in part funded by the U.S. Department of Energy under grant DE-SC0024532. MAF thanks Corpus Christi College, Cambridge and the Cambridge Trust for a studentship, as well as Peterhouse for funding through a Research Fellowship. This work used the ARCHER2 UK National Supercomputing Service (https://www.archer2.ac.uk).

%%%END OF MAIN TEXT%%%


\balance


%%%REFERENCES%%%
\bibliography{references} %You need to replace "rsc" on this line with the name of your .bib file
\bibliographystyle{rsc} %the RSC's .bst file
\appendix

\section{Higher Taylor expansions of the exponential projector}
\label{appendix:no_benefit}
\begin{figure}[!htp]
    \centering
    \includegraphics[width=\columnwidth]{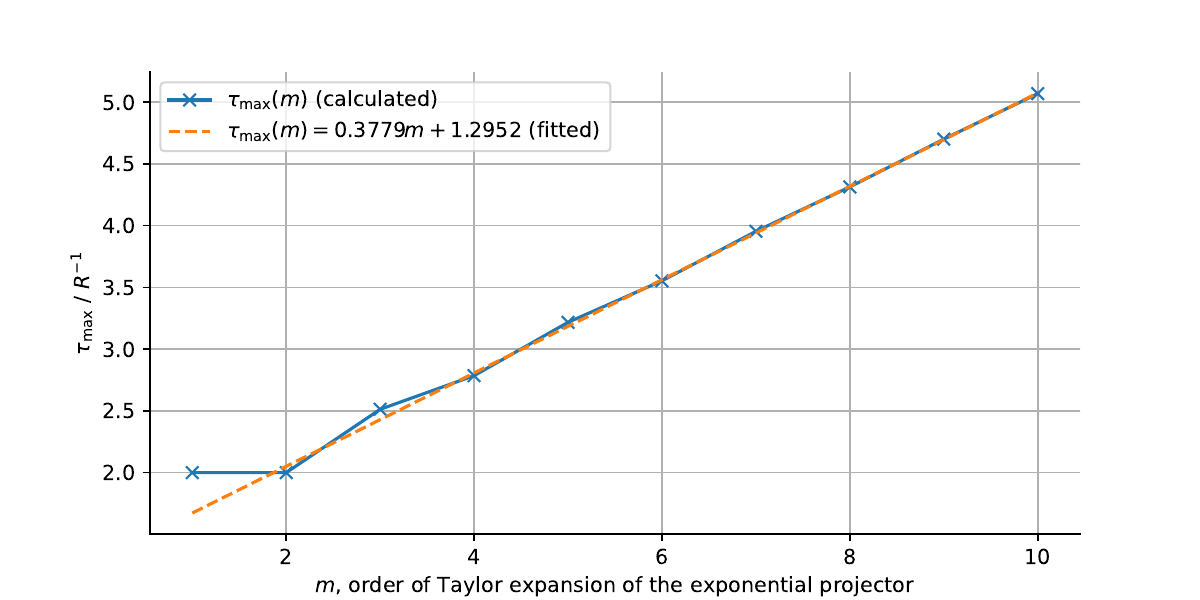}
    \caption{Calculated and fitted $\tau_{\text{max}}$ as a function of the order of Taylor expansion of the exponential projector. There is indeed no gain whatsoever in going to the second order Taylor expansion, but there is in going to yet higher orders.}
    \label{fig:tau_max}
\end{figure}

In Ref. \citenum{zhang_2016} it was proposed that there is no gain in going to higher order Taylor expansions of the exponential projector, because all orders of expansion have $\gamma=\tau$ (see \Cref{appendix:conv}). The conclusion is correct, but for a more subtle reason that we will now explain. The $m$-th order Taylor series expansion of $g^{\text{exp}}$ is
\begin{equation}
    \sum_{k=0}^m\frac{1}{k!}(-\tau)^k(x-E_0)^k. 
\end{equation}
Eq. 6 in Ref. \citenum{zhang_2016} requires that $g(E_{N-1})<1$, so, defining the \textit{spectral range} $R=E_{N-1}-E_0$, we have
\begin{equation}
\label{eq:max_tau}
    \left|\sum_{k=0}^m\frac{1}{k!}(-\tau)^kR^k \right|<1.
\end{equation}
The $m=1$ case leads to the familiar requirement in DMC, FCIQMC and CCMC that
\begin{equation}
    \tau_{\text{max}} <\frac{2}{E_{{N-1}}-E_0}
\end{equation}
and solving \Cref{eq:max_tau} numerically shows that higher order expansions lead to larger maximum allowed $\tau$. In fact, asymptotically, $\tau_{\text{max}}$ increases linearly with a gradient of $1/e\approx0.368$ (see \Cref{fig:tau_max}). So, na\"ively we can expect the efficiency to increase linearly with $m$. To prove the linearity, we note that $\tau_{\text{max}}R>1$, and we can approximate $\sum_{k=0}^{m}|(-x)^k/{k!}|$ ($x\equiv\tau R$) with the leading order term $|(-x)^m|/{m!}$, and so we are left with
\begin{subequations}
\begin{align}
|(-x)^m|&<m!\\
m\ln x&\lessapprox m\ln m-m\\
x&<me^{-1},
\end{align}
\end{subequations}
where we used the Stirling's formula in the second line. 
Therein lies the real reason for not using higher order expansions: a naive implementation requires $m(m+1)/2$ applications of the Hamiltonian per projection, and even a factorised implementation would require $m$ applications per projection, not to mention the lack of closed forms for the roots of the $m$-th order expansion, due to the Abel-Ruffini theorem\cite{ayoub_paolo_1980}, which shows that no analytical solution can exist for $m\geq5$. In any case, the overall efficiency stays at best constant. Therefore, the conclusion that no gains can be made is correct, although a more tortuous argument is needed.

\section{Properties of the wall-Chebyshev projector}
\label{appendix:wall-ch}
Assuming the entire spectral range is re-scaled such that $x\in [-1,1]$, where $x = 2(E-E_0))/R - 1 $, and $R=E_{N-1}-E_0$ is the \textit{spectral range} or the Hamiltonian, the Chebyshev expansion coefficients of the wall function is
\begin{equation}
\begin{split}
c_k=&\int_{-1}^1 \text{wall}(x+1)T_k(x) (1-x^2)^{-1/2} \dl x\\
\equiv&\int^1_{-1}\text{wall}(x+1)T_k(x)\delta(x+1) \dl x \\
=&(2-\delta_{k0})T_k(-1)=(2-\delta_{k0})(-1)^k,
\end{split}
\end{equation}
where the second line uses the fact that the wall function is zero everywhere but at the lower bound, so the weight function $(1-x^2)^{-1/2}$ has the same action as the delta function centred at $-1$, $\delta(x+1)$. The last equality exploits a well-known identity of the Chebyshev polynomials \cite{riley_hobson_bence_2006}. We can then write the $m$-th order expansion as
\begin{equation}
\label{eq:appendix-wall-ch}
\begin{split}
    g^{\text{wall-Ch}}_m(x)&= \frac{1}{1+2m}\sum_{k=0}^m (2-\delta_{k0}) (-1)^kT_k(x)\\
    &=\frac{1}{1+2m}\sum_{k=0}^m(2-\delta_{k0})T_k(-x),
\end{split}
\end{equation}
where the last equality exploits the fact that $T_k$ has the same parity as $k$ \cite{riley_hobson_bence_2006}, and we scale the sum such that $g^{\text{wall-Ch}}_m(-1)=1$. 

To obtain an analytical expression for the zeroes of the wall-Chebyshev projector, we use the trigonometric definition of the Chebyshev polynomials. Inverting the sign of the argument in \Cref{eq:appendix-wall-ch}, we have
\begin{equation}
\label{eq:appendix-wall-ch-coeff}
\begin{split}
    g^{\text{wall-Ch}}_m(-\cos\theta)&\propto \sum_{k=0}^m(2-\delta_{k0})T_k(\cos\theta)\\
    &=1+2\left[\sum_{k=0}^m\cos(k\theta) \right]\\
    &=\frac{\sin\left[{\left(m+1/2\right)\theta}\right]}{\sin\left(\theta/2\right)},
\end{split}
\end{equation}
where the last equality is the Dirichlet kernel \cite{lejeune1829convergence}. The zeroes of $g^{\text{wall-Ch}}_m(x)$ are then transparently
\begin{equation}
    a_{\nu}=-\cos\left(\frac{\nu\pi}{m+1/2} \right),\ \nu=1,2,\dots,m,
\end{equation}
where the negative sign is from account for the sign inversion in \Cref{eq:appendix-wall-ch-coeff}. In an arbitrary spectral range other than $[-1,1]$, these zeroes are
\begin{equation}
    a_{\nu}=E_0+\frac{R}{2}\left(1-\cos\frac{\nu\pi}{m+1/2} \right).
\end{equation}
Knowing its zeroes, we can decompose $g^{\text{wall-Ch}}_m(x)$ into a product of linear projectors:
\begin{equation}
    g^{\text{wall-Ch}}_m(x) = \prod_{\nu=1}^m \frac{x-a_{\nu}}{E_0-a_{\nu}},
\end{equation}
where the numerators ensure $g^{\text{wall-Ch}}_m(E_0)=1$. 

\section{Convergence properties of generators}
\label{appendix:conv}
We summarise here some important properties of generators. The asymptotic rate of convergence of a propagator is dominated by the slowest-decaying eigencomponent:
\begin{equation}
    \mu = \lim_{n\rightarrow\infty}=\frac{||\Psi^{(n+1)}-\Psi_0 ||}{||\Psi^{(n)}-\Psi_0 ||} = \max_i \left| \frac{g(E_i)}{g(E_0)} \right|.
\end{equation}
Zhang and Evangelista \cite{zhang_2016} suggested that, in the common case that the first excited state is the slowest-decaying component, and that the first excited energy is small compared to the spectral range of $\hat{H}$, the above can be approximated as 
\begin{equation}
    \mu\approx \left|1+(E_1-E_0)g'(E_0) \right|\equiv\left|1-\alpha\gamma \right|, 
\end{equation}
where $\gamma=-g'(E_0) $ is the \textit{convergence factor} for $g$. We now derive the relation given in Ref. \citenum{kosloff_direct_1986} that $\gamma$ is approximately the number of times, $n$, that $g$ needs to be applied to achieve an error in the norm, $\epsilon=||\Psi^{(n)}-\Psi_0||$ to the $N$-th decimal place:
\begin{subequations}
 \begin{align}
 \epsilon=10^{-N}&\approx (1-\alpha\gamma)^n\\
10^{-N}&\approx e^{-n\alpha\gamma}\\
n&=\frac{N\ln10}{\alpha\gamma}\equiv\kappa N \cdot \frac{1}{\gamma},
 \end{align}
 \end{subequations} 
where $\kappa=\ln10/(E_1-E_0)$ is the convergence prefactor, which is inversely proportional to the first excited energy gap.
\end{document}